\documentclass[aps,prl,preprint,groupedaddress,showpacs]{revtex4}
\usepackage{graphicx}

\begin{document}


\title{NEW ENHANCED TUNNELING IN NUCLEAR PROCESSES}


\author{Boris Ivlev }
\affiliation{Department of Physics and Astronomy\\
University of South Carolina, Columbia, SC 29208\\
and \\
Instituto de F\'{\i}sica, Universidad Aut\'onoma de San Luis Potos\'{\i}\\
San Luis Potos\'{\i}, S. L. P. 78000 Mexico}
\author{Vladimir Gudkov }
\email[]{gudkov@sc.edu}
\affiliation{Department of Physics and Astronomy\\
University of South Carolina, Columbia, SC 29208}



\begin{abstract}
The small sub-barrier tunneling probability of nuclear processes can be dramatically enhanced by collision with incident charged particles. Semiclassical methods of theory of complex trajectories have been applied to nuclear tunneling, and conditions for the effects have been  obtained. We demonstrate the enhancement of alpha particle decay by incident proton with energy of about $0.25\hspace{0.1cm}{\rm MeV}$. We 
show that the general features of this process are common for other sub-barrier nuclear processes and can be applied to nuclear fission. 
\end{abstract}


\maketitle



Tunneling in nuclear processes has been a subject of study for many years since this is a substantial mechanism of nuclear decay and nuclear reactions, including nuclear fission and fusion (see, for example, \cite{fis,fus} and references therein). The recent interest in understanding the processes of under-barrier tunneling \cite{kasagi,bertsch,takigawa,Zel1,Zel2,Dyak2} has been stimulated by the calculation of bremsstrahlung radiation in alpha decay when the alpha particle is moving under the barrier \cite{Dyak1}. In this letter we consider another feature of quantum tunneling: the possible enhancement of nuclear decay due to interactions with low energy charged particles. This enhancement has the same origin as the tunneling enhancement in nonstationary fields recently discovered in condensed matter physics \cite{IVLEV1,IVLEV2,IVLEV3}, and can manifest itself in different under-barrier processes. 

For the sake of simplicity, we  consider nuclear alpha decay. According to the  theory of Gamov, the probability of the transition of alpha particle through the nuclear Coulomb barrier is mainly ruled by the exponential factor \cite{WEISSKOPF,BOHR}
\begin{equation}
\label{prob}
W\sim\exp\left[-A_{\alpha}(E)\right],
\end{equation}
where
\begin{equation}
\label{act0}
A_{\alpha}(E)=\frac{\sqrt{8M}}{\hbar}\int^{R_{\alpha}}_{R_{0}}dR\hspace{0.1cm}\sqrt{\frac{2Ze^{2}}{R}-E}
\end{equation}
is the classical action measured in units of $\hbar$ \cite{LANDAU}. $M$ is the mass of alpha particle, $Z$ is the charge of the daughter nucleus, $R_{0}$ is the nuclear radius, and the classical exit point $R_{\alpha}$ is determined by zero of the square root. Let us study how this probability changes when the decayed nuclei are placed in the beam of protons with the energy less than the barrier height $2Ze^{2}/R_{0}$. This situation rersembles the process of quantum tunneling of particles controlled by a weak and varying in time electromagnetic field considered in \cite{MELN1,MELN2,MELN3,IVLEV1,IVLEV2,IVLEV3}, where the specific tunneling enhancement mechanisms have been studied. The difference between our case and these processes is mainly in the nature of the external electromagnetic field (the beam of protons, in our case).  The low energy projectile protons can be treated as a source of a pulsed electromagnetic fields interacting with the alpha decayed nuclear target. 

According to the results of papers \cite{IVLEV1,IVLEV2,IVLEV3}, two different regimes of tunneling are possible if the proton energy and its time of the under-barrier motion satisfy the necessary conditions. The first regime is the assistance of tunneling, when the alpha particle gains a part of proton energy, which can be called the positive assistance. The second one, which occurs when the alpha particle transfers a part of its energy to the proton, is called the negative assistance of tunneling. Under conditions of the negative assistance, the alpha particle tunnels at lower energy where the barrier is less transparent. Nevertheless, contrary  to any expectation, the regime of negative assistance of tunneling is unusual, since, under certain conditions, tunneling probability does not become exponentially small even for a barrier which is normally not very transparent. This phenomenon is called Euclidean resonance. Both mechanisms of positive and negative assistance are connected with the coherent multiquanta interference in the under-barrier motion. We know that the enhancement of tunneling occurs when a singularity of the nonstationary field coincides in position  with a singularity of the classical Newtonian trajectory of the particle on the complex time plane. To study these processes in nuclei, we apply the semiclassical approach (based on the method of trajectories in complex time) developed for tunneling in  nonstationary fields. 

Consider the assistance of alpha decay by the Coulomb field of incident protons when its energy is less than the height of the Coulomb barrier. In the absence of a proton, one can calculate the decay probability using formulas (\ref{prob}) and (\ref{act0}).  A low energy incoming proton interacts with the nucleus only electromagnetically, since it is stopped by the nuclear Coulomb field at a distance much larger than the radius of strong nuclear forces. The time of interactions of the proton with the decayed  alpha particle (under-barrier motion time) is about of a characteristic nuclear time $\Delta t\sim 10^{-21}{\rm s}$. The proton can excite the nucleus to increase the energy of emitted alpha particle $E$ by the amount of $\Delta E$ during the interaction time $\Delta t$. This leads to the increase of the action in Eq.~(\ref{prob}) by the amount $2\Delta t\Delta E/\hbar$ (the rigorous definition of $\Delta t$ is given below). At the same time, the energy gain by the excited alpha particle makes its tunneling easier due to the reduction of the action $A_{\alpha}(E)\rightarrow A_{\alpha}(E+\Delta E)$. Then, the resulting action for the proton induced alpha decay is 
\begin{equation}
\label{actal}
A=\frac{2 \Delta t\Delta E}{\hbar}+A_{\alpha}(E+\Delta E).
\end{equation}
In condensed matter physics, this equation would describe the process of the positive photon-assisted tunneling with the probability $\exp(-A)$ \cite{IVLEV1}. The first term in (\ref{actal}) results in a reduction of the probability due to quanta absorption and the second one describes tunneling in a more transparent (higher energy) part of the barrier. Since the flux of tunneling particles in a nonstationary field is also a nonstationary one, the expression $\exp(-A)$ relates to the maximal value of the tunneling probability. This maximal transition probability is determined by a finite value of the energy transfer $\Delta E$, which provides a minimum of the action $A$, and, hence, it is defined by the condition $\partial A(E+\Delta E)/\partial\Delta E = 0$. 

The Eq.~(\ref{actal}) describes the tunneling assisted by a given external nonstationary field. In our case, the role of this field is played by the proton not with  a ``given'' but a rather with a ``flexible'' motion affected by both the nucleus and the alpha particle. The increase of the value of alpha particle energy must be accompanied by the corresponding decrease of the proton energy value (from its initial value $\varepsilon$ down to $(\varepsilon - \Delta E)$). This means that the alpha particle and the proton participate in a cooperative tunneling motion from the nuclear surface  to the outside of the barrier region which can be  described by the joint action $A_{\alpha p}$. However, since the proton does not tunnel through the Coulomb barrier, the true action should be corrected as\cite{IVLEV3} 
\begin{equation}
\label{10}
A=A_{\alpha p}-\frac{2i\sigma_{p}}{\hbar},
\end{equation}
where the second term (being real and negative) accounts for the phase shift of the process of the  artificial move of the proton from outside of the barrier to the nuclear surface. The classical imaginary action of the proton $\sigma_{p}$  can be determined from the corresponding Hamilton-Jacobi equation. At the limit of a weak alpha-proton Coulomb coupling, one can estimate $2i\sigma_{p}/\hbar\simeq A_{p}(\varepsilon)$, where $A_{p}(\varepsilon)$ is the proton analog of the action (\ref{act0})
\begin{equation}
\label{11}
A_{p}(\varepsilon)=\frac{\sqrt{8m}}{\hbar}\int^{r_{e}}_{R_{0}}dr\hspace{0.1cm}\sqrt{\frac{Z_{0}e^{2}}{r}-\varepsilon}.
\end{equation}
Here, $m$ is the proton mass and $Z_{0}$ is the charge of decayed nucleus. The  action $A_{\alpha p}$ is defined on the joint alpha-proton classical trajectory in imaginary time when both particles, with initial energies $E$ (alpha particle) and $\varepsilon$ (proton), meet on the nuclear surface. The energy exchange between the alpha particle and the proton occurs fast and weakly contributes  to the joint action $A_{\alpha p}$ which can be written as $A_{\alpha p}\simeq A_{\alpha}(E+\Delta E)+A_{p}(\varepsilon -\Delta E)$. Then, the total action takes the form
\begin{equation}
\label{acttot}
A=A_{\alpha}(E+\Delta E)+A_{p}(\varepsilon-\Delta E)-A_{p}(\varepsilon).
\end{equation}
The classical time $\tau_{0}$ of the under-barrier motion, which is proportional to the derivative of the action with respect to energy, has  the same value both for the alpha particle and for the proton since they move together
\begin{equation}
\label{12}
\frac{2\tau_{0}}{\hbar}=-\frac{\partial A_{\alpha}(E+\Delta E)}{\partial\Delta E}=
\frac{\partial A_{p}(\varepsilon -\Delta E)}{\partial\Delta E}.
\end{equation}
Therefore,  Eq.~(\ref{12}) determines the certain energy transfer $\Delta E$ which provides an extreme of the action (\ref{acttot}). At the limit of a very small energy transfer ($\Delta E\ll\varepsilon$),  the proton motion becomes  non-``flexible''. This corresponds to the action of Eq.~(\ref{actal}), with $\Delta t=\tau_{0}$, which can be obtained from  Eq.~(\ref{acttot}) after expansion on $\Delta E$. In this case, the tunneling motion of the alpha particle is affected by the nonstationary field
\begin{equation}
\label{15}
V_{int}(\vec{R},i\tau)=\frac{2e^2}{|\vec{R}-\vec{r}(i\tau)|},
\end{equation}
where $\vec{r}(i\tau)$  approximately describes the classical trajectory of the free proton. The total action (\ref{acttot}) can be written in the 
explicit form  
\begin{equation}
\label{13}
A=\frac{2{\pi}Ze^2}{\hbar}\sqrt{\frac{2M}{E+\Delta E}}-4\hspace{0.1cm}\sqrt{\frac{R_{0}}{\hbar^{2}/(4MZe^2)}}+
\frac{\pi Z_{0}e^2\sqrt{2m}}{\hbar}\left(\frac{1}{\sqrt{\varepsilon -\Delta E}}-\frac{1}{\sqrt{\varepsilon}}\right)  .       
\end{equation}
Then, the relation between the optimum energy transfer $\Delta E$ and the energy of the incident proton $\varepsilon$ is given by  Eq.~(\ref{12}):
\begin{equation}
\label{14}
\frac{\varepsilon -\Delta E}{E+\Delta E}=\left(\frac{m }{4M}\right)^{1/3},
\end{equation}
where we disregard the small difference between charges of the parent and daughter nuclei. It should be noted that generally the energy transfer $\Delta E$ is determined by  Newtonian equations for the alpha particle and the proton in imaginary time, which are coupled by the interaction (\ref{15}). As a consequence, the value of $\Delta E$ depends on the angle $\phi$ between directions of radial motions of these two particles (we consider  both particles to have zero angular momenta). For example, $\Delta E$ is positive for $\phi =180\hspace{0.04cm}^{\circ}$ and negative for $\phi =0\hspace{0.04cm}^{\circ}$ (parallel motion). This means that the condition (\ref{14}) of the optimum energy transfer is fulfilled for a certain angle between directions of classical motion of two particles. 

One can see that  the energy transfer $\Delta E$ during the tunneling process can be either positive (positive assistance of tunneling) or negative (negative assistance of tunneling). The latter case, as  mentioned above, is unusual, since the action can tend to zero and tunneling probability does not become exponentially small (Euclidean resonance) even for a barrier which is normally hardly transparent. Indeed, by substituting  expression (\ref{14}) into Eq.~(\ref{13}) we obtain 
\begin{equation}
\label{17}
A=\frac{2\pi Z_{0}e^{2}}{\hbar}\sqrt{\frac{2M}{E+\varepsilon}}\left[1+\left(\frac{m}{4M}\right)^{1/3}\right]^{3/2}-
\frac{8}{\hbar}\sqrt{MZ_{0}e^{2}R_{0}}-\frac{\pi Z_{0}e^{2}}{\hbar}\sqrt{\frac{2m}{\varepsilon}}.
\end{equation}
If $\Delta E$ is negative, the proton energy $\varepsilon$ can be chosen small (see Eq.~(\ref{14})) and the last term in Eq.~(\ref{17}) may reduce the action $A$ down to a zero value. It should be noted that the above equation  is correct if $\exp(-A)\ll 1$. When $A$ becomes of the order of unity or less, one should use a generic formalism  with the multi-instanton approach, which leads to the similar estimate of the action $A$.

Let us give an example of the calculation of the energy transfer $\Delta E$  using the method of complex trajectories. Consider a classical parallel motion of the alpha particle and the proton (the angle between particles $\phi =0$) when only $x$-components are involved and are determined by the Newton equations
\begin{equation}
\label{61}
M\hspace{0.1cm}\frac{\partial R_{x}}{\partial\tau^{2}}=-\frac{2Ze^{2}}{R^{2}_{x}}+\frac{2e^{2}}{(r_{x}-R_{x})^{2}}
\hspace{0.1cm};
\hspace{1cm}m\hspace{0.1cm}\frac{\partial r_{x}}{\partial\tau^{2}}=-\frac{Z_{0}e^{2}}{r^{2}_{x}}-\frac{2e^{2}}{(r_{x}-R_{x})^{2}}
\end{equation}
In the vicinity of the complex time $\tau_{0}$, when particles meet each other,  the solutions of these equations have the form
\begin{equation}
\label{62}
\frac{R_{x}(i\tau)}{R_{s}}=\frac{r_{x}(i\tau)}{r_{s}}=\left(\frac{\tau_{0}-\tau}{\tau_{0}}\right)^{2/3}
\end{equation}
where $R_{s}$ and $r_{s}$ are some constants to be defined. The energy $\Delta E$, gained by the alpha particle, 
\begin{equation}
\label{63}
\Delta E=2e^{2}\int^{\tau_{0}}_{0}\frac{d\tau}{(R_{x}-r_{x})^{2}}\hspace{0.1cm}\frac{\partial R_{x}}{\partial\tau}
\end{equation}     
diverges close to the $\tau_{0}$ and should be regularized by the condition $R_{s}(1-\tau/\tau_{0})^{2/3}>R_{0}$. (It should be noted, that in contrast to the large contribution of the diverged interaction $V_{int}$ to $\Delta E$, its contribution to the action $A$ is not divergent and even rather small.) Then,
\begin{equation}
\label{64}
|\Delta E|=\frac{2e^{2}}{R_{0}}\left(\frac{r_{s}}{R_{s}}-1\right)^{-2},
\end{equation} 
where the ratio $R_{s}/r_{s}$ satisfies the relation
\begin{equation}
\label{65}
\frac{M}{2m}\left(\frac{R_{s}}{r_{s}}\right)^{3}
\left[\left(1-\frac{R_{s}}{r_{s}}\right)^{2}+\frac{2}{Z_{0}}\right]=
\left(1-\frac{R_{s}}{r_{s}}\right)^{2}-\frac{1}{Z_{0}}\left(\frac{R_{s}}{r_{s}}\right)^{2}. 
\end{equation}
Considering the uraniun alpha decay as an example 
\begin{equation}
\label{33}
^{235}_{92}{\rm U}+p\rightarrow\hspace{0.1cm}^{231}_{90}{\rm Th}+\alpha + p
\end{equation}
with the energy  of alpha particle $E=4.678$ Mev, we can fix  the ratio $M/m=4$, and  obtain the parameter $R_{s}/r_{s}\simeq 0.715$. These lead to the energy transfer $|\Delta E|\simeq 1.89$\hspace{0.1cm}MeV and  non-physical (negative) value of $\varepsilon$. This means that in the case of the classical trajectory with $\phi =0$, the energy transfer is larger than the optimum value (see Eq.~(\ref{14})). The optimum $\Delta E$ (which leads to an extreme value of the action $A$ ) corresponds to a finite $\phi$ of the classical trajectory and can be found numerically using the  above scheme. 

The analysis of the expression for the action (\ref{17}) shows that at $\varepsilon =\varepsilon_{0}$, where
\begin{equation}
\label{67}
\varepsilon_{0}=E\left(\frac{m}{4M}\right)^{1/3}\simeq 1.85\hspace{0.1cm}{\rm MeV},
\end{equation}
 the energy transfer $\Delta E=0$, when the  angle $\phi\simeq 30{\hspace{0.02cm}}^{\circ}$. Under those conditions, the action (\ref{17}) coincides with  the action of the conventional alpha decay (\ref{act0}), resulting in the tunneling probability
\begin{equation}
\label{68}
W\sim\exp\left[-A_{\alpha}(4.678\hspace{0.1cm}{\rm MeV})\right]\simeq e^{-80.75}\simeq 10^{-35}
\end{equation}
The probability (\ref{68}), being normalized by  the nuclear attempt frequency $10^{21}\hspace{0.1cm}{\rm s}^{-1}$, describes experimental data reasonably well \cite{WEISSKOPF,BOHR}. 

\begin{figure}
\includegraphics[width=6in]{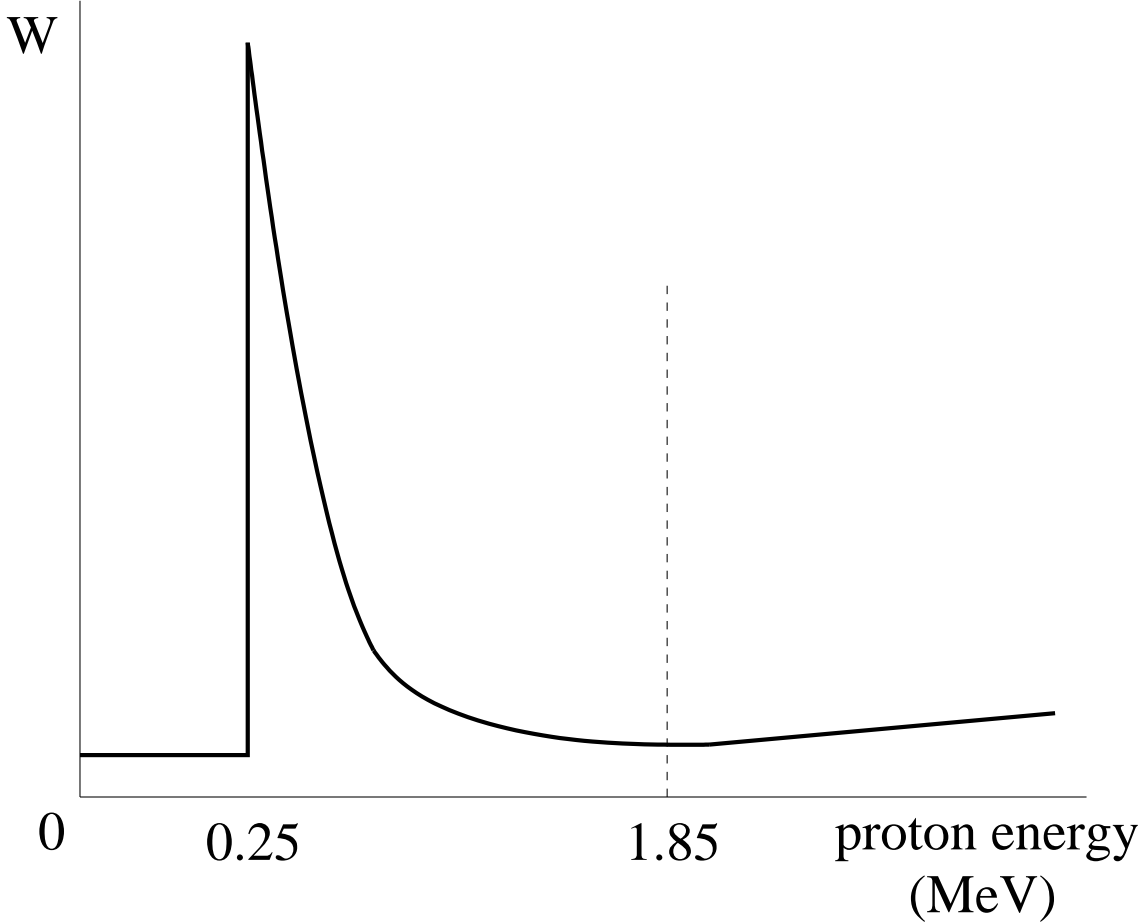}
\caption{\label{fig1}
The right side of the energy dependence (smooth enhancement) of the tunneling probability corresponds to the domain $\varepsilon >1.85$\hspace{0.1cm}MeV and is related to  positive assistance of tunneling. The Euclidean resonance occurs at $\varepsilon =0.25$\hspace{0.1cm}MeV  at the region of negative assistance tunneling $\varepsilon <1.85$\hspace{0.1cm}MeV.}
\end{figure}

At $\varepsilon <\varepsilon_{0}$, the optimum energy transfer $\Delta E$ becomes finite and negative, the optimum angle $\phi$ decreases, and the action (\ref{17}) reduces in comparison with $A_{\alpha}(E)$. Upon reduction of $\varepsilon$, the action (\ref{17}) turns to zero at a certain proton energy $\varepsilon_{R}$, which relates to Euclidean resonance. For the reaction (\ref{33}) $\varepsilon_{R}=0.25$\hspace{0.1cm}MeV, the accompanied energy transfer is $\Delta E=-1.15$\hspace{0.1cm}MeV, and the angle between the incident proton and the emitted alpha particle is $\phi\simeq 11{\hspace{0.02cm}}^\circ$. In other words, when  the energy of the incident proton is about $\varepsilon =0.25$\hspace{0.1cm}MeV, the energy of the emitted alpha particle  becomes $E-|\Delta E| =3.53$\hspace{0.1cm}MeV (instead of $E=4.678$\hspace{0.1cm}MeV), and the energy of outgoing proton becomes  $\varepsilon +|\Delta E| =1.40$\hspace{0.1cm}MeV. The cross-section of the reaction (\ref{33}) is not exponentially small at $\varepsilon =\varepsilon_{R}$ and has a very sharp peak in the vicinity of the proton energy. The typical behavior of the tunneling probability as a function of the proton energy $\varepsilon$ is shown in Fig.~\ref{fig1}. 
Positive assistance of tunneling ($\Delta E>0$) corresponds to the domain $\varepsilon >1.85$\hspace{0.1cm}MeV and Euclidean resonance at $\varepsilon =0.25$\hspace{0.1cm}MeV occurs at the region of negative assistance $\varepsilon <1.85$\hspace{0.1cm}MeV. At $\varepsilon>\varepsilon_{R}$, the shape of the peak is proportional to $\exp[-(\varepsilon-\varepsilon_{R})/\Delta\varepsilon]$, where
$\Delta\varepsilon\simeq 10^{-4}$MeV.  

\begin{figure}
\includegraphics[width=16cm]{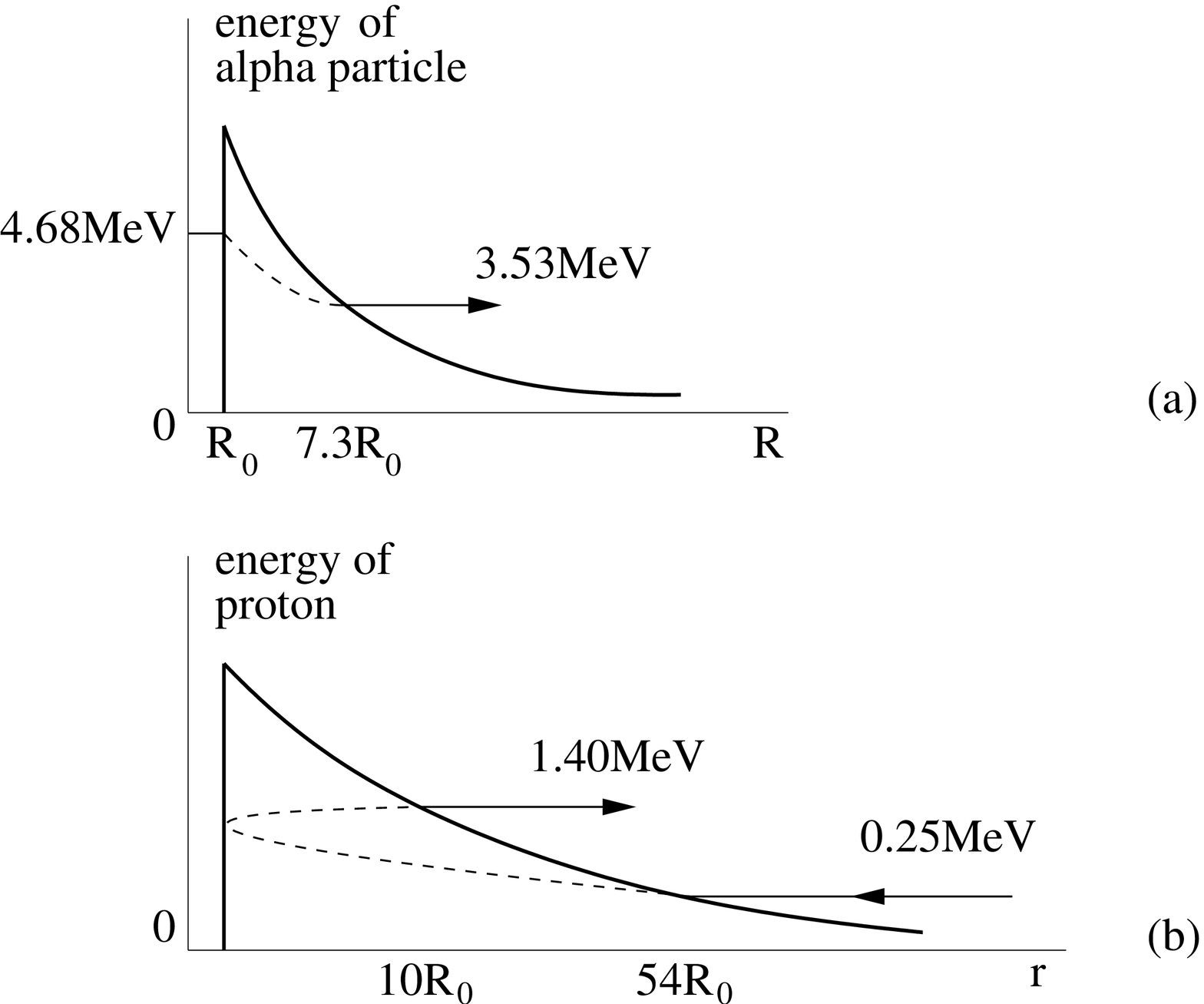}
\caption{\label{fig2}(a) The trajectory of the alpha particle; (b) The trajectory of the proton. These classical trajectories in imaginary time are only a convenient way to describe the effect. In real time the proton does not approach the nucleus and interacts with it solely via the Coulomb field.}
\end{figure}

It should be noted that at $\varepsilon =\varepsilon_{R}$, the exit point of alpha particle is $7.3R_{0}$ and  of the outgoing the proton is $10R_{0}$. The incident proton is stopped by the nuclear Coulomb field at the distance of about $54R_{0}$, as shown in Fig.~\ref{fig2}.
Therefore, the  trajectories in imaginary time are only a convenient language to describe the effect. In real (physical) time, the proton does not approach the nucleus and interacts with it solely via the Coulomb field. 

 Within the exponential accuracy, the tunneling probability can be calculated  by solving either the static problem or the equivalent dynamical one with the nonstationary field (\ref{15}). As it is known, the {\it normal} tunneling through a  static barrier can be described simply by a classical trajectory in imaginary time, which connects two classically allowed regions. This mechanism has always been employed in the study of  tunneling \cite{OKUN,STONE,COLEMAN,MELN4}. In the language of nonstationary field, the normal tunneling corresponds to the case when the tunneling particle does not absorb quanta of the nonstationary field \cite{IVLEV3}. In addition to this process, there is an {\it enhanced} tunneling through a  barrier which corresponds, in the language of nonstationary field, to the absorption or emission of quanta (positive or negative assistance) \cite{IVLEV3}. The description of the enhanced tunneling, in contrast to the normal one, does not reduce only to the classical trajectory and a narrow bundle of paths in its vicinity but also requires one to take into account the delicate interference of various paths. This can be seen from Eq.~(\ref{10}), where the second part of the right side of the expression is not originated form a classical trajectory. In other words, the enhanced tunneling is a coherent multi-dimensional tunneling.  It should be noted that we use the method of complex trajectories for calculations of under-barrier processes instead of numerical solution   of the Schr\"{o}dinger equation mostly because of an insufficiency of the existing algorithms to solve  this type of Schr\"{o}dinger equation in a reasonable amount of time (see, also ref. \cite{Zel1}).

The idea of stimulation of nuclear tunneling processes by incident charged particles can be applied in the similar way to nuclear fission processes  on the basis of models with nuclear fragments tunneling under the action of the external varying Coulomb field. However, the validity of this approach for fusion reactions is less obvious and requires a further study.

In  summary, protons, approaching alpha decaying nuclei,  create the nonstationary Coulomb field, acting on the tunneling alpha particle. Due to these interactions the Euclidean resonance can appear at low proton energy and the Coulomb barrier becomes practically transparent for the  passage of the alpha particle. For example, normally, $^{235}_{92}$U emits alpha particle with the energy of  $4.678$\hspace{0.1cm}MeV. When the energy of the incident proton is close to its resonant value of $0.25$\hspace{0.1cm}MeV, the energy of outgoing protons becomes $1.40$\hspace{0.1cm}MeV and the energy of emitted alpha particles becomes   $3.53$\hspace{0.1cm}MeV.

\end{document}